\documentclass[reprint,superscriptaddress,amsmath,amssymb,aps,pre]{revtex4-1}
\usepackage{graphicx}
\usepackage{caption}
\usepackage{overpic}
\usepackage{dcolumn}
\usepackage{bm}
\usepackage{subfigure}
\usepackage{amsmath}
\usepackage{amsfonts}
\usepackage{color}
\usepackage{array} 
\usepackage{booktabs} %
\usepackage{rotating}
\usepackage[english]{babel}
\usepackage{amsthm,amsmath,amssymb}
\usepackage{mathrsfs}
\usepackage{lipsum}
\usepackage{dutchcal}
\usepackage{float}
\usepackage{changes}
\usepackage{mathptmx}
\DeclareMathAlphabet{\mathcal}{OMS}{cmsy}{m}{n}
\DeclareSymbolFont{largesymbols}{OMX}{cmex}{m}{n}
\let\sum\relax
\DeclareSymbolFont{CMlargesymbols}{OMX}{cmex}{m}{n}
\DeclareMathSymbol{\sum}{\mathop}{CMlargesymbols}{"50}

\usepackage{booktabs,array,float,tabularx,booktabs,lipsum,amsmath,multirow}
\usepackage[hidelinks]{hyperref}
\hypersetup{
    colorlinks=true,
    linkcolor=blue,
    filecolor=magenta,
    urlcolor=blue,
    citecolor=magenta,
    }   

\begin{document}

\title{Pulse-driven depinning of magnetic gap modes in ferromagnetic films}

\author{Xin-Wei Jin}
\affiliation{School of Physics, Northwest University, Xi'an 710127, China}
\affiliation{Peng Huanwu Center for Fundamental Theory, Xi'an 710127, China}

\author{Zhan-Ying Yang}
\email{zyyang@nwu.edu.cn}
\affiliation{School of Physics, Northwest University, Xi'an 710127, China}
\affiliation{Peng Huanwu Center for Fundamental Theory, Xi'an 710127, China}

\author{Yanan Liu}
\email{yanan.liu@nwu.edu.cn}
\affiliation{School of Physics, Northwest University, Xi'an 710127, China}

\author{Guangyin Jing}
\affiliation{School of Physics, Northwest University, Xi'an 710127, China}

\date{\today}

\begin{abstract}

Manipulation of magnons in artificial magnonic crystals (MCs) leads to fascinating nonlinear wave phenomena such as the generation of gap solitons, which has been mostly limited to one-dimensional systems.
Here, we propose a model system for the magnetization in two-dimensional MCs subjected to a periodic external magnetic field, describing the dynamics of magnetic gap solitons (MGSs) formed by nonlinear self-trapping. We show the formation, stability, and dynamics for various two-dimensional gap modes, including gap solitons and vortical ones. Their existence regions depend on the anisotropic axis orientation of the ferromagnetic film. The Bloch oscillation and depinning propagation of MGSs under constant spin-current injections are discovered and characterized. We design a scheme of pulse current injection to  achieve distortionless propagation of MGSs. These findings show that the 2D magnonic crystals can be viewed as a building block for MGSs-based storage and transmission, where the propagation and localization are variously controlled and reconfigurable.

\end{abstract}

\maketitle

\section{Introduction}

Periodic lattice modulation, enabling the generation of reconfigurable band structures, is recognized as a versatile and convenient tool for manipulating wave dispersion, offering novel avenues for wave control within various nonlinear systems \cite{kartashov2011solitons, fleischer2003observation, zeng2020preventing, kartashov2013gap, yang2023spin}.
This mechanism hold great significance both for their achievable applications such as light deceleration \cite{john2012trap} and signal storage-recovery \cite{chumak2012storage} and for their importance in fundamental research in nonlinear physics \cite{chen1993gap}. Various interesting localization phenomena, like gap solitons and truncated Bloch waves, can occur due to the interplay between the band structure and nonlinear interactions \cite{tanese2013polariton, kivshar1993self, pernet2022gap, zhang2009gap, kim2017chiral}.

Among various artificial crystal systems, magnonic crystals (MCs) attract growing attentions \cite{puszkarski2003magnonic, wang2009observation, wang2010nanostructured, qin2019experimental}.
These arise from the wide-range methods for creating periodic magnetic structures, including  macroscopic metallic stripes or dots, etched grooves or pits, periodic external magnetic field and periodic saturation magnetization by ion implantation \cite{krawczyk2014review, duerr2012enhanced, tacchi2012mode}.
Such rich external potential structure and the possibility of fast dynamic control of magnon via the spin-transfer torque (STT) and spin-orbit torque (SOT) effects \cite{zhang2004roles, li2004domain2, miron2011perpendicular, liu2012spin, dohi2019formation} make MCs promising candidates for the fundamental studies as well as for the magnetic storage and transfer application. It has also been shown that the shape and strength of the pinning potential and the driving force can strongly affect the soliton dynamics \cite{li2021robust}. In general, designs from periodic structures by taking advantage of material nature or external electromagnetic fields, can be abstracted as the introduced external potentials into the systems. These additional potentions are expected to manipulate the transport behavior of quasi-particles or wave-modes transport, and are able to provide essential artificial pinning for precise control of gap soliton positions.

Futhermore, considerable theoretical and experimental efforts have been devoted to investigate different localization modes in MCs \cite{ustinov2010formation, morozova2016band, morozova2022nonlinear}. Up to now, a large part of investigations of localization modes in MCs is concentrated in one-dimensional system \cite{morozova2016band, morozova2022nonlinear, morozova2012mechanisms, wu2008coupled, morozova2022gap, langer2019spin}. While a few of reports on two-dimensional MCs is limited to experimental detection of spin wave dispersion curves and the design of diverse band structures \cite{tacchi2011band, tacchi2012forbidden, krawczyk2013magnonic, chen2022magic}.
However, some exciting two-dimensional localized gap modes were predicted and observed in optical and cold atomic system \cite{fleischer2003observation, pedri2005two}. Therefore, the magnetic localization modes in higher-dimensional lattices are expected to present qualitatively distinct features since the symmetry and dimensionality of the lattice begin to play a crucial role in the formation and characteristics of band structures and their corresponding nonlinear modes \cite{ostrovskaya2003matter}. Many fundamental characteristics are anticipated to emerge in higher dimensions, such as lattice vortex solitons and gap solitons carrying angular momentum, analogous to those that occur in photonic lattices and Bose-Einstein condensates \cite{fleischer2003observation, lobanov2014fundamental, huang2021fundamental, xu2023vortex}. As for 2D MCs, the existence, stability, and dynamics of localized gap modes have yet to be unveiled. In addtion, investigating the driving effect of STT on MGSs is also demanding in the context of magnetic storage applications.

In this paper, we propose an effective model for the magnetization in two-dimensional MCs subjected to a periodic external magnetic field, describing the dynamics of 2D self-trapping magnetic gap modes. Our investigation unveils that these 2D localized gap modes manifest as magnetic gap solitons and magnetic vortices, residing exclusively within the band gaps. The existence region of these magnetic gap solitons depends entirely on the anisotropic axis of the ferromagnetic film. By employing a combination of linear stability analysis and direct numerical simulations, the stability regions of MGSs are determined. Under a constant spin-current excitation, we show the MGSs undergo spatial oscillations within the initial lattice. This behavior can be understood as Bloch oscillation in the effective potential well. Upon elevated spin-current injection, MGSs commence depinning, manifesting deformation during diffusion. We finally introduce a scheme of pulse current injection to overcome such distortion and achieve distortionless propagation of MGSs. These findings underscore the potential for generating and controlling GSs within 2D MCs.

\section{Modeling}\label{SecII}

\begin{figure}[ht]
\vspace{0cm} %
\centering
\includegraphics[width=8.5cm]{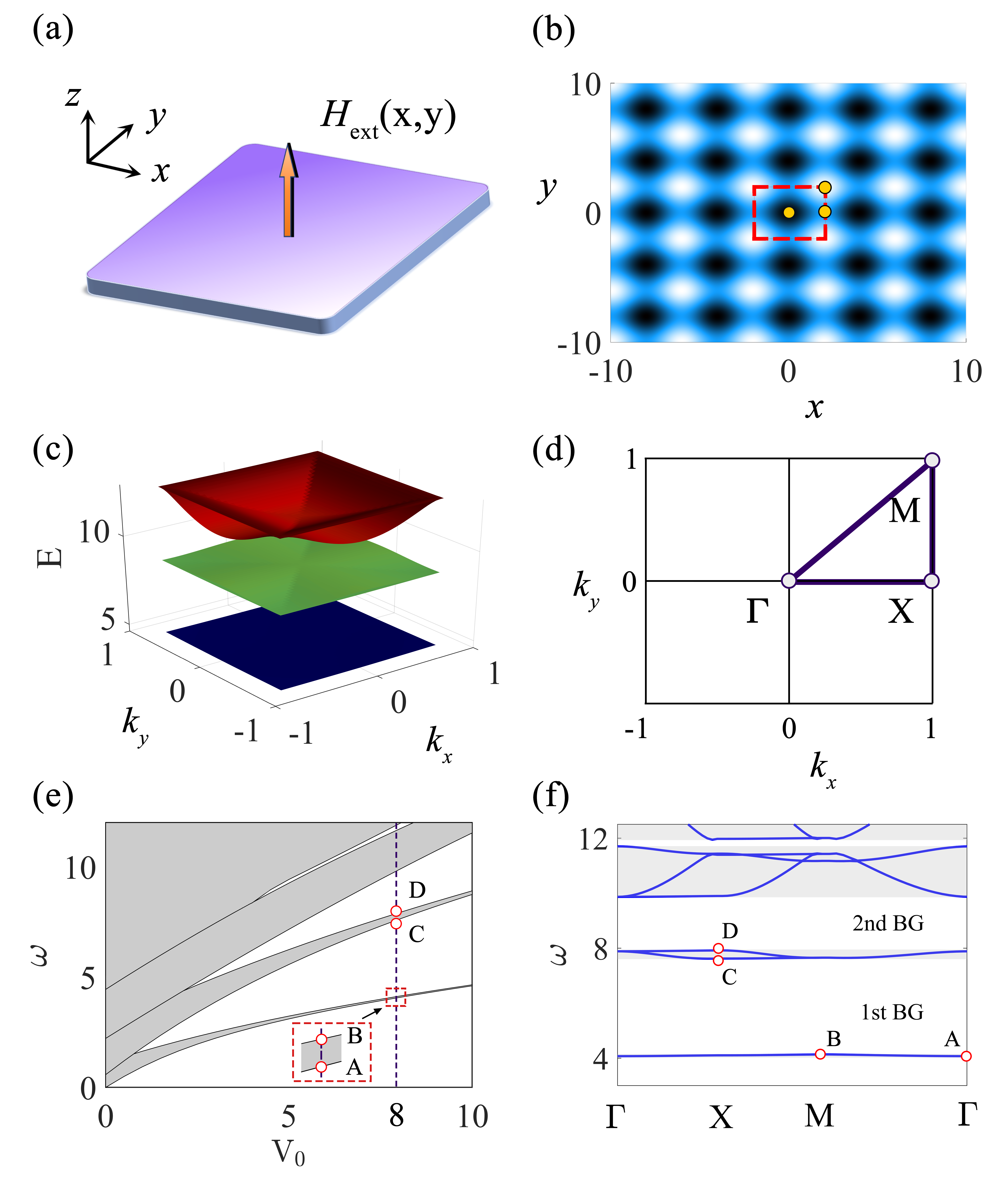}
\caption{Sketch of the 2D MCs and corresponding band structure. (a) Ferromagnetic film under periodic external magnetic field $\textbf{H}_{\rm{eff}}$. (b) Density plot of the square external magnetic field $\textbf{H}_{\rm{eff}}=H_{0}\left[\sin(\pi x/a)^{2}+\sin(\pi y/a)^{2}\right]$. (c) Dispersion surfaces of the linear Eq. (\ref{sGP}) at $H_{0}=8$. (d) The first reduced Brillouin zone of the high symmetry points in reciprocal lattice space. (e) Band gap structure for the Bloch waves with different $H_{0}$. The gray area and white area respectively represent the energy band and bandgap, and the letters $A, B, C, D$ mark the band edges at $H_{0}=8$. (f) The dispersion spectrum at $H_{0}=8$.}\label{Schem}
\end{figure}

An uniaxial anisotropic magnetic thin film under the external magnetic field $H_{\rm{ext}}$, is considered as shown in Fig. \ref{Schem}(a).
The dynamics of magnetization $\textbf{M}(x, y, t)$ is governed by the Landau-Lifshitz-Gilbert (LLG) equation \cite{li2004domain2, landau1965collected, yuan2022quantum}:
\begin{equation}\label{LLG}
\frac{\partial\textbf{M}}{\partial t}=-\gamma\textbf{M}\times\textbf{H}_{\rm{eff}}+\frac{\alpha}{M_{s}}\left(\textbf{M}\times\frac{\partial\textbf{M}}{\partial t}\right)+{\bm \tau}_{b},
\end{equation}
where $M_{s}$ is the saturation magnetization, $\gamma$ and $\alpha$ are the gyromagnetic constant and the Gilbert damping parameter, respectively.
The effective field $\textbf{H}_{\rm{eff}}$ is given by
$
\textbf{H}_{\rm{eff}}=H_{\rm{ext}}+\frac{2A}{M_{s}^{2}}\nabla^{2}\textbf{M}+\frac{2K_{u}}{M_{s}}(\textbf{M}\cdot\textbf{n})\textbf{n},
$
where $H_{\rm{ext}}$ denotes the external magnetic field, $A$ is the exchange constant, $K_{u}$ is the anisotropy constant, and $\mathbf{n}=(0,0,1)$ is the unit vector directed along the anisotropy axis. The ferromagnetic films are referred to as easy-axis magnets when $K_{u}>0$, and as easy-plane magnets when $K_{u}<0$ \cite{iacocca2017breaking, li2004domain2}.
The last term $\tau_{b}$ on the right-hand side of Eq. (\ref{LLG}) represents some applied torques such as STT and SOT. We now consider a spin-current injection along the $x$-direction, therefore the corresponding adiabatic STT can be written as
$
\tau_{b}=-\frac{b_{J}}{M_{s}^{2}}\textbf{M}\times(\textbf{M}\times\frac{\partial\textbf{M}}{\partial x})
$.
Here, we have defined the effective driving velocity $b_{J}=Pj_{e}\mu_{B}/eM_{s}$, in which $P$ is the spin polarization of the current, $j_{e}$ is the electric current density, $\mu_{B}$ is the Bohr magneton, and $e$ is the magnitude of electron charge \cite{li2004domain2}.
Accounting for the restriction $\textbf{M}^{2}=M_{s}^{2}$, we introduce a complex variable $\Psi(x,y,t)=(M^{x}+iM^{y})/M_{s}$. When considering a small deviation of magnetization from the equilibrium direction, corresponding to $\lvert\Psi\rvert^{2}\ll1$, it is reasonable to obtain $M^{z}/M_{s}\approx1-\lvert\Psi\rvert^{2}/2$.
Substituting this ansatz into Eq. (\ref{LLG}), neglecting damping, and further introducing a gauge transformation $\Psi\to e^{2iK_{u}}\Psi$, we obtain the effective magnetization dynamical equation in the presence of STT:
\begin{equation}\label{GPE}
i\frac{\partial\Psi}{\partial t}=2A\nabla^{2}\Psi+K_{u}\lvert\Psi\rvert^{2}\Psi-H_{\rm ext}\Psi+ib_{J}\Psi.
\end{equation}
Herein the external magnetic field $H_{\rm{ext}}$ can be regarded as an equivalent external potential, while the sign of $K_{u}$ determines distinct nonlinear types.
We assume a square magnetic trap [whose contour plot is shown in Fig. \ref{Schem}(b)] in the form
$
H_{\rm{ext}}=H_{0}\left[\sin(\pi x/a)^{2}+\sin(\pi y/a)^{2}\right],
$
where $H_{0}$ and $a$ characterize the modulation amplitude and period of the magnonic crystal.

\begin{figure*}[ht]
\vspace{0cm} %
\centering
\includegraphics[width=15.0cm]{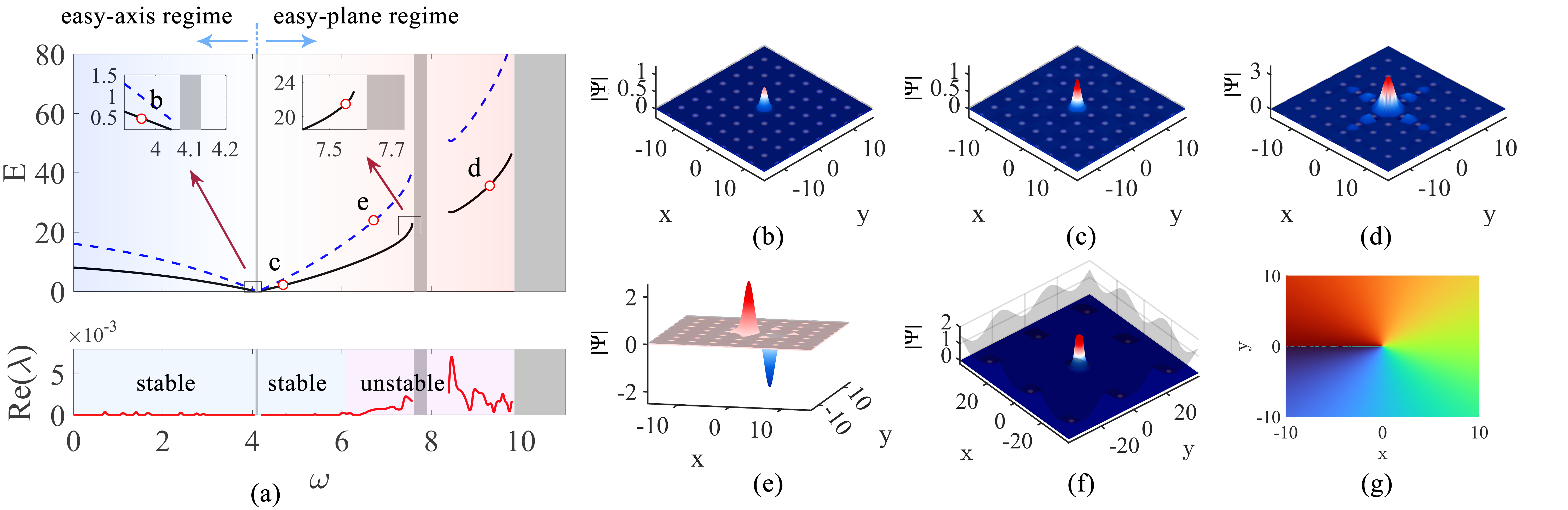}
\caption{Stability property of 2D localized gap modes denoted as the polarization energy $E$ versus propagation constant $\omega$. (a) Upper panel: existence curves of fundamental MGSs (black solid line) and dipole MGSs (blue dashed line); Bottom panel: maximum real part of linear spectrum for fundamental MGSs. Characteristic shapes of MGSs at different propagation constant $\omega$: (b) $\omega$=3.9; (c) $\omega$=4.2; (d) $\omega$=9.5; (e) $\omega$=7.55; (f)(g) Characteristic shape and phase distribution of gap vortex soliton.}\label{phase}
\end{figure*}

\section{Results}\label{SecIII}
\subsection{Bloch modes and band structure}

To investigate the existence of 2D nonlinear localized gap modes, it is essential to explore the band structure that hold pivotal significance within a periodic medium.
For the linearized model of Eq. (\ref{GPE}), the steady solitary solutions are shown in the form $\Psi(x,y,t)=\psi(x,y)e^{-i\omega t}$, with $\omega$ being magnetization energy of the system.
It thus becomes
\begin{equation}\label{sGP}
\omega\psi-2A\nabla^{2}\psi+H_{ext}\psi-ib_{J}\psi=0.
\end{equation}
The stationary solutions are found by applying Bloch's theorem, which states that the wave function has the form $\psi(x,y)=e^{ik_{x}x+ik_{y}y}\phi(x,y)$, with $k_{x}$ and $k_{y}$ being wavenumbers in the first Brillouin zone $-1\le k_{x},k_{y} \le 1$. The functions $\phi(x,y)$ are periodic in the $x$-direction and $y$-direction of period $a$, i.e. $\phi(x+a,y)=\phi(x,y+a)=\phi(x,y)$.

For the given sinusoidal square magnetic lattice, the guided Bloch modes and dispersion relation can be computed effectively by a Fourier collocation method \cite{yang2010nonlinear}.
As a result, the 2D band structure (dispersion surfaces) with $H_{0}=8$ and $a=4$ is shown in Fig. \ref{Schem}(c). From the illustration, it is evident that two wide band gaps appear between the lowest three dispersion surfaces.
Fig. \ref{Schem}(d) depict the first Brillouin zone and high-symmetry points of the reciprocal lattice space.
At other values of the magnetic field strength $H_{0}$, we have summarized the dispersion diagram of Eq. (\ref{sGP}) in Fig. \ref{Schem}(e).
This plot reveals the emergence of finite band gaps at nonzero magnetic field strength $H_{0}$, with both the number and width of these band gaps expanding as $H_{0}$ increases.
Fig. \ref{Schem}(f) shows the band structure at magnetic field strength $H_{0}=8$.
Particularly, the edges of the Bloch bands are respectively marked in Fig. \ref{Schem}(e) by red hollow circles and labeled with the letters A, B, C, and D, in alignment with the annotations in Fig. \ref{Schem}(f).

\subsection{Families of localized gap modes and corresponding stability}

After successfully identifying the finite band gaps within the linear spectrum of the model (\ref{GPE}), we now turn to explore the formation of 2D nonlinear localized gap modes. Generally these gap modes emerge as a consequence of the interplay between the given periodic magnetic lattice and cubic non-linearities.
Hence, their characteristics and existence regions are intimately tied to the sign of the nonlinear terms, which, in turn, corresponds to two distinct anisotropy type of the ferromagnetic thin film.

\begin{figure}[b]
\vspace{0cm} %
\centering
\includegraphics[width=7.2cm]{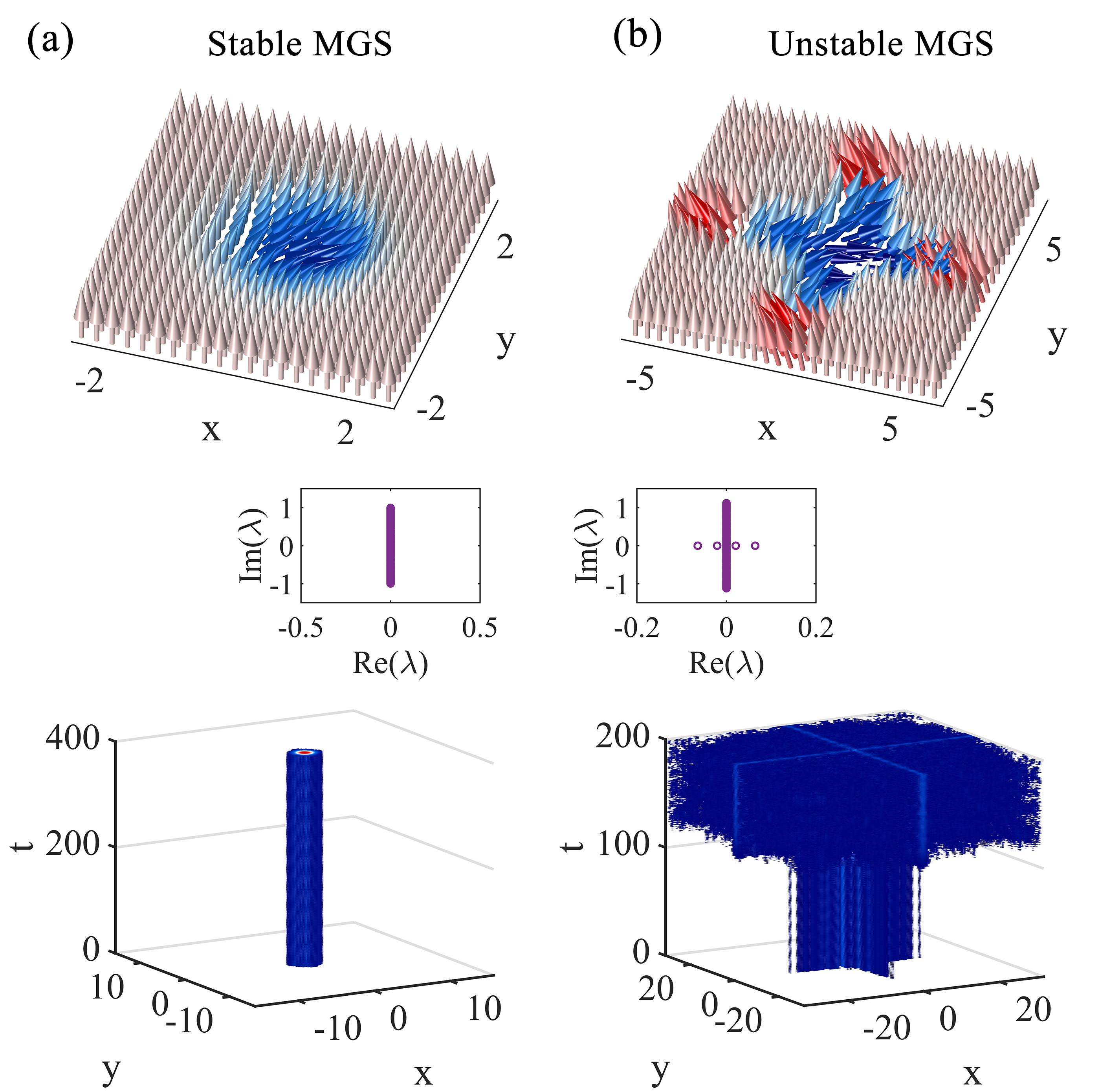}
\caption{Characteristic stability property of a stable MGS and an unstable MGS. (a) Magnetization vector distribution of stable MGS at propagation constant $\omega=4.5$. Perturbed evolution and the corresponding whole stability spectrum. (b) Magnetization vector distribution of unstable MGS at propagation constant $\omega=6.8$. Perturbed evolution and the corresponding whole stability spectrum.}\label{example}
\end{figure}

\begin{figure*}[ht]
\vspace{0cm} %
\centering
\includegraphics[width=14cm]{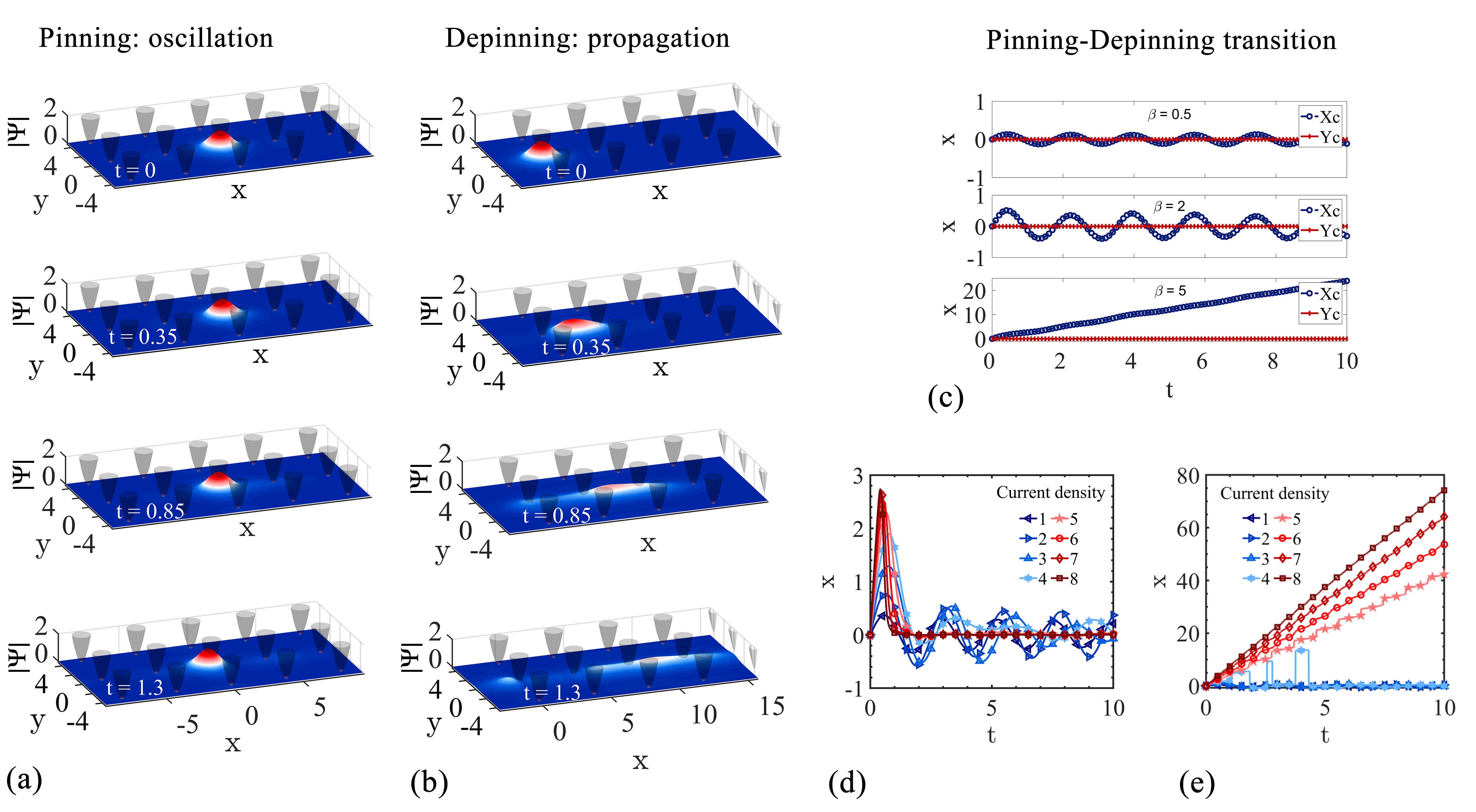}
\caption{(a) Snapshots of bloch oscillation of MGS under low constant spin-current excitation $b_{J}=1$. (b) Snapshots of diffuse propagation of MGS under high constant spin-current excitation $b_{J}=8$. (c) Movements of the magnetic center of MGSs under three intensities of spin currents. (d) Magnetization center motion of MGS in the initial Wigner-Seitz cell under different spin currents. (e) Evolution of the maximum position of MGS under different spin currents.}\label{pdp}
\end{figure*}

The steady state of a gap soliton in both easy-axis and easy-plane regime Eq. (\ref{sGP}) are numerically solved by the Newton conjugate-gradient method.
The initial wave packet of $\psi(x,y)$ is assumed as a 2D Gaussian wave packet $\psi(x,y)=\beta e^{-(x^{2}+y^{2})/2\sigma^{2}}$ with an amplitude $\beta$ and width $\sigma$ on the scale of several lattice constants. We note that spin current injection is not taken into consideration here, and its effect will be discussed in next subsection.
Our numerical results are shown in Fig. \ref{phase}.
Among the interesting features of this phase diagram is that, the existence regions of MGSs are completely separated by the first energy band. In the case of an easy-plane ferromagnetic thin film ($K_{u}< 0$), MGSs only reside within the semi-infinite band gap, which corresponds to the left portion of Fig. \ref{phase}(a). In contrast, for easy-axis ferromagnetic thin films, MGSs corresponding to these films exclusively reside in the first and second band gaps (right portion of Fig. \ref{phase}(a)).
In both existence regions, two classes of soliton families including fundamental, dipole MGSs and vortex solitons are identified.
The polarization curve of MGS families are displayed on Fig. \ref{phase}(a), denoted by the black solid and blue dashed lines.
Here the polarization energy is defined as $E=\iint \psi{\rm d}x{\rm d}y$.

Our calculation unveils various families of interesting 2D gap modes and here we showcase some categories. Fig. \ref{phase}(b)-(d) depict three characteristic fundamental MGSs with a single main peak, located in the first and second band gaps, corresponding to magnetization energy $\omega = 3.9, 4.5$ and $9.5$, respectively. By comparing the profiles of fundamental MGSs within the same family, it can be observed that as $\omega$ moves away from the edge of first energy band, the amplitudes or total polarizations of the MGSs gradually increase, accompanied by the emergence of nearby modulation structures. The profile of a dipole MGS is depicted in Fig. \ref{phase}(e), residing on the first band gap.
Fig. \ref{phase}(f) and \ref{phase}(g) present the profile and phase of a gap vortex mode with topological charge $Q=1$.

While the periodic external ``effective potential" aids in stabilizing two-dimensional solitons, the stability of these 2D gap solitons remains an important issue to be determined. We employ linear stability analysis and perturbation propagation techniques to validate the stability of fundamental gap solitons in the semi-infinite band, and those within the first and second band gaps.
To analyze the linear stability of this solitary wave, we consider it is perturbed by normal modes as $\widetilde{\psi}=\left\{\psi+[v(x,y)+w(x,y)]e^{\lambda t}+[v^{*}(x,y)-w^{*}(x,y)]e^{\lambda^{*} t}\right\}e^{i\omega\tau}$,
here $\psi=\psi(x,y)$ represents the unperturbed wave function of Eq. (\ref{sGP}), $v(x,y)$ and $w(x,y)$ are small perturbations for a given eigenvalue $\lambda$.
Substituting this perturbed solution into Eq. (\ref{sGP}) and linearizing thereafter, we obtain the following linear-stability eigenvalue problem:
\begin{equation}\label{C2}
\begin{pmatrix}
G_{0} & \nabla^{2}+G_{1} \\
\nabla^{2}+G_{2} & -G_{0}
\end{pmatrix}
\begin{pmatrix}
v \\
w \\
\end{pmatrix}
=
-i\lambda
\begin{pmatrix}
v \\
w
\end{pmatrix},
\end{equation}
with $G_{0}=K_{u}(\psi^{*2}-\psi^{2})/2-ib_{J},\
G_{1}=-\omega-2K_{u}|\psi|^2+K_{u}(\psi^{2}+\psi^{*2})/2+H_{\rm ext},\
G_{2}=-\omega-2K_{u}|\psi|^2-K_{u}(\psi^{2}+\psi^{*2})/2+H_{\rm ext}.$ In general, the above eigenvalue problem can be efficiently solved using the Fourier collocation method, allowing us to obtain the corresponding eigenvalues $\lambda$ and assess the stability conditions of the perturbed 2D gap soliton.

The numerical results are presented in the bottom panel of Fig. \ref{phase}(a) by the ${\rm Re}(\lambda) \sim \omega$ curve. Each point on the curve represents the maximum real part of the linear spectrum for the corresponding MGS. From the curve, it reveals that all MGSs within the semi-infinite bands are linear stable, corresponding to the easy-plane regime (where $K_{u} < 0$).
In the easy-axis regime ($K_{u} > 0$), part of MGSs in the first band gap are linear stable, indicated by the light blue region in the figure. As the magnetization energy of the MGSs in this band gap increases, their total polarization also increases, and instability begins to emerge. The total polarization in the second band gap continues to rise, and at this point, all MGSs are unstable.

The representative profiles, eigenvalues and perturbed evolutions of both stable and unstable modes are shown in Fig. \ref{example}. We show the vector magnetization distribution of the stable MGS Fig. \ref{example}(a), corresponding to point $c$ in Fig. \ref{phase}(a). The linear eigenvalue spectrum is present in the upper right corner of Fig. \ref{example}(a), exactly matching the direct perturbation evolution in Figure \ref{example}(c). Similarly, Figs. \ref{example}(b) and \ref{example}(d) present stability tests for the unstable MGS at point $d$ in the second band gap of Fig. \ref{phase}(a).
The direct perturbation evolution in Fig. \ref{example}(d) indicates weak instability near $t=100$ under the influence of small perturbations, accompanied by rapid changes in the global configuration. This result also consists with predictions from the related linear eigenvalue spectrum.

\subsection{Pinning-depinning transition for current-driven gap soliton}

To achieve reliable control of the soliton position in spintronic devices, external torques are often introduced by spin-current injections. It has been show that the adiabatic spin-transfer torque can be
utilized to depin the DW. In this subsection, we will show the driven motion of the MGS under the influence of STT. First, let us adopt a small spin-current injection with drive parameter $b_{J}=1$. In this case, the MGS cannot overcome the effective magnetic potential barrier in the initial Wigner-Seitz (WS) cell.
The Bloch oscillation of MGS is consequently excited. Fig. \ref{pdp}(a) shows the oscillatory behavior of the MGS within the WS cell, with an oscillation period of approximately $t = 1.3$. It is conceivable that as the spin current increases, the MGS will accumulate enough energy to enable it to traverse the equivalent magnetic potential barrier, achieving inter-cell transport. We also present the snapshots of MGS dynamics for a large drive parameter $b_{J}=5$ in Fig. \ref{pdp}(b).
For such drives above a critical value, the oscillating solution is no longer possible, and the MGS is entirely activated, evolves dynamically to adjacent ``equivalent potential well" at constant drive.

To quantitatively measure the motion of MGS under the influence of different spin-currents, we introduce a magnetization center (or equivalent centroid), which is calculated as:
\begin{equation}\label{XY}
X_{c}=\frac{\iint_{-\infty}^{+\infty} x\ \psi {\rm d}x{\rm d}y}{\iint_{-\infty}^{+\infty} \psi {\rm d}x{\rm d}y}, \ \ Y_{c}=\frac{\iint_{-\infty}^{+\infty} y\ \psi {\rm d}x{\rm d}y}{\iint_{-\infty}^{+\infty} \psi {\rm d}x{\rm d}y},
\end{equation}
where $X_c$ and $Y_c$ represent the components of the magnetic center in the $x$ and $y$ directions, respectively. Through the magnetic center, we can effectively characterize the transition of MGSs from pinning to depinning as the external spin current increases. Fig. \ref{pdp}(c) illustrates the movement of the magnetic center of MGSs under different spin currents. It is worth noting that the spin current is applied along the $x$-axis, hence the MGSs are subjected to a ``driving force" solely in the $x$-axis direction. As seen in the subfigures, there are no transversal movements in the center of the MGS. For small driving parameter $b_J=0.5$, the MGS undergoes minor oscillations within the initial WS cell. As the driving parameter $b_J$ increases to 2, the MGS remains unable to overcome the magnetic potential barrier, resulting in periodic oscillations with a significant increase in oscillation displacement amplitude. When the driving parameter $b_J$ is set to 5, the MGS depins and is driven in the positive $x$-axis direction.

Two additional effective pinning parameters are introduced to investigate the transitional region: (i) the coordinates of the magnetization center in the initial WS cell, calculated by
$X_{ws}=({\iint_{-a/2}^{+a/2} x\ \psi {\rm d}x{\rm d}y})/({\iint_{-a/2}^{+a/2} \psi {\rm d}x{\rm d}y}),$ $Y_{ws}=({\iint_{-a/2}^{+a/2} y\ \psi {\rm d}x{\rm d}y})/({\iint_{-a/2}^{+a/2} \psi {\rm d}x{\rm d}y});$
and (ii) the position of the MGS's maximum amplitude $X_{m}$ and $Y_{m}$, which aids in characterizing the trajectory of the MGS as it moves, excluding the dispersed impurity waves. All these quantities are numerically calculated from the simulations.
The pinning parameters versus time with various current densities in plotted in Fig. \ref{pdp}(d) and \ref{pdp}(e), from which the pinning-depinning transition region can be simply determined.
From the results it is evident that when the spin current density $b_{J}<3$, the MGS remains pinned within the initial cell.
While the MGS resides in the depinning region when the spin current density $b_{J}>4$.
Above the transition region, the magnetization center of initial WS cell rapidly decays to nearly zero, indicating they are in the depinning state. The evolution of the maximum position of the MGS is shown in Fig. \ref{pdp}(d), which also confirm the pinning-depinning transition.

\subsection{Spin-current pulse-driven perfect migration}

\begin{figure}[t]
\vspace{0cm} %
\centering
\includegraphics[width=7.2cm]{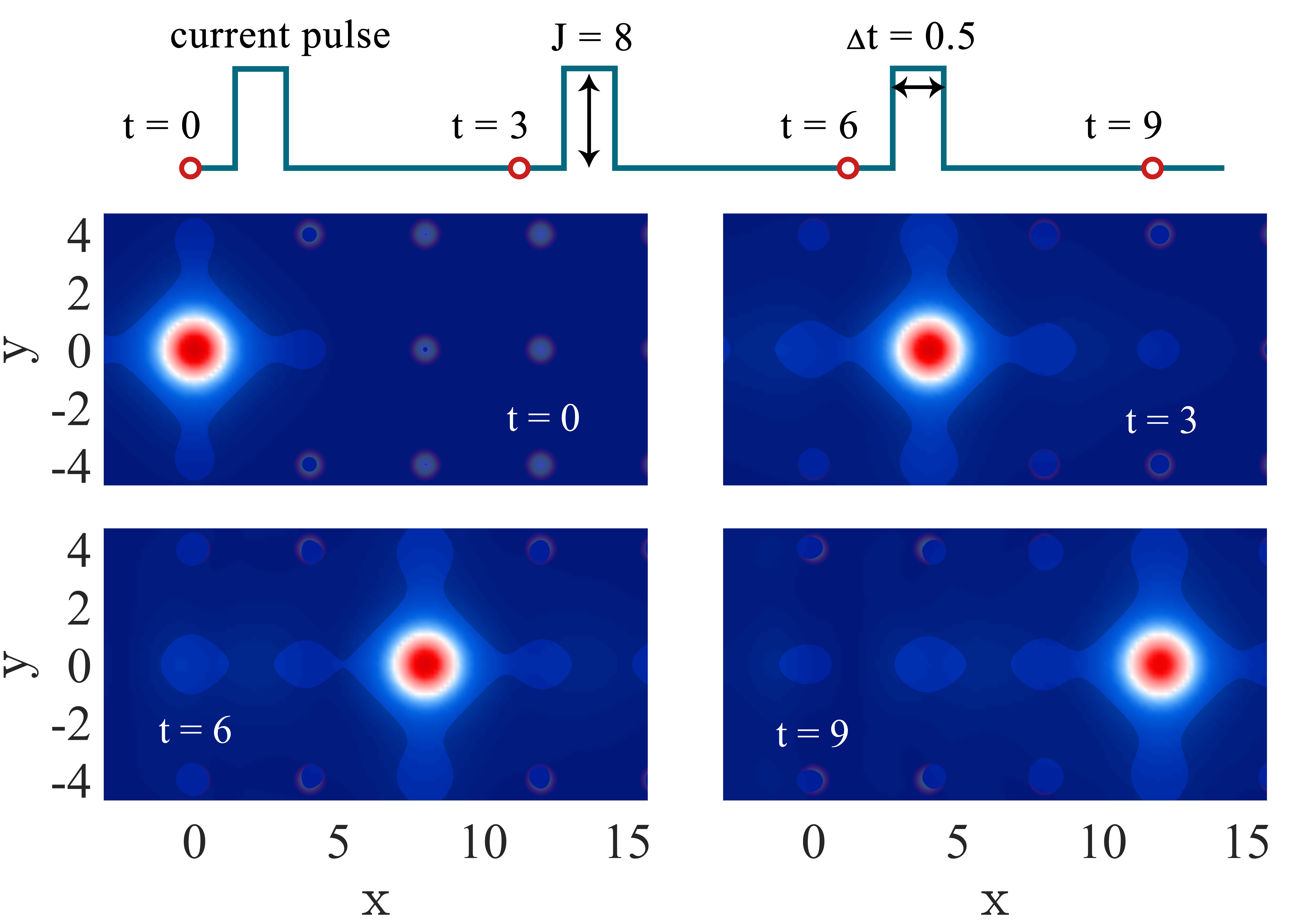}
\caption{Snapshots of the distortionless propagation of MGS achieved by the pulse-current scheme.}\label{pulse}
\end{figure}

An important observation from the above driven motion is that, in the presence of a large constant driving current, MGSs undergo deformation during their transport across potential barriers. This can be understood by the differential response of magnetization spin configurations with varying orientations to the spin current, resulting in a diffuse propagation of MGS along the direction of the spin current. Such a dispersion process is detrimental to applications in information storage based on MGSs. To achieve the overall migration of MGSs between potential wells, we introduce a relaxation time after crossing the ``effective potential barriers". The perfect overall migration precession of the MGS under spin current pulse is given in Fig. \ref{pulse}.
As shown in the top panel, the square pulse's intensity is configured with $b_{J}=8$ and a pulse width of $\Delta t=0.5$. Consequently, the MGS precisely shifts to the next ``effective potential well" upon receiving each spin current pulse. The time gap between two pulses is $\Delta t=2.5$, ensuring that the MGS has sufficient time for recovering to its initial profile.
The snapshots of MGS dynamics at $t=0,3,6$ and 9 are shown in the bottom four panels of Fig. \ref{pulse}.

\section{Discussion and conclusion}\label{SecIV}

In conclusion, we propose a quantitative and numerical study of  the formation, stability, and dynamics for
two-dimensional gap modes in ferromagnetic films. In the presence of two-dimensional periodic external magnetic field, an effective model are derived to describe the MGS dynamics. The linear band structure are obtained from the Bloch approach. We predicted the existence of MGSs in both easy-axis and easy-plane ferromagnetic films. It is of interest to find a clear distinction in the regions where MGSs exist in easy-axis and easy-plane ferromagnetic films. They are respectively located within the band gaps on either side of the first energy band. Combining the linear stability analysis with direct perturbed simulations of the dynamical equations, we have determined the stable and unstable regions for all localized gap modes.

Considering the spin-current injection, we investigate the MGS pinning dynamics at different current densities. Both the numerical simulation and analytical theory show that the adiabatic STT acts as a driving force on the MGS. The Bloch oscillation and depinning propagation of MGS under constant spin-current injections are discovered and characterized. In the scenario of large spin-current injection, MGSs undergo deformation and even fragmentation while being driven, attributed to the asynchrony in the velocities of different magnetizations crossing the equivalent barriers. We propose a scheme of pulse current injection to achieve distortionless propagation of MGSs. These results provide insight into a fundamental micromagnetic process that could be useful for current-driven magnetic storage and magnonic devices based on MGSs.

\section*{Acknowledgment}

We appreciate fruitful discussions with Prof. Haiming Yu and Prof. Zhimin Liao. This work was supported by the National Natural Science Foundation of China (Nos. 12275213, 12174306, 12247103), Natural Science Basic Research Program of Shaanxi (2023-JC-JQ-02, 2021JCW-19) and 2023 Graduate Innovation Program of Northwest University No. CX2023100.

\bibliography{references}
\end{document}